# Quantifying memories: mapping urban perception

**Shan He[a], Yuji Yoshimura[b], Jonas Helfer[c], Gary Hack[d], Carlo Ratti[b], Takehiko Nagakura[e]**

[a] Uber Technologies Inc, 1455 Market St, San Francisco, CA 94103-1355, USA;

[b] SENSEable City Laboratory, Massachusetts Institute of Technology, 77 Massachusetts Avenue, Cambridge, MA 02139, USA;

[c] Databricks, 160 Spear Street, 13th Floor, San Francisco, CA 94105, USA;

[d] Department of Urban Studies and Planning, Massachusetts Institute of Technology, 77 Massachusetts Avenue, Cambridge, MA 02139, USA;

[e] Department of Architecture, Massachusetts Institute of Technology, 77 Massachusetts Avenue, Cambridge, MA 02139, USA.

**Abstract**
What people choose to see, like, or remember is of profound interest to city planners and architects. Previous research suggests what people are more likely to store in their memory – buildings with dominant shapes and bright colors, historical sites, and intruding signs - yet little has been done by the systematic survey. This paper attempts to understand the relationships between the spatial structure of the built environment and inhabitants' memory of the city derived from their perceptual knowledge. For this purpose, we employed the web-based visual survey in the form of a geo-guessing game. This enables us to externalize people's spatial knowledge as a large-scale dataset. The result sheds light on unknown aspects of the cognitive role in exploring the built environment, and hidden patterns embedded in the relationship between the spatial elements and the mental map.

**Keywords**
City image, cognitive mapping, urban computing, visual perception

## 1. Introduction

Lynch (1960)[1] argues the image of the city that people navigate in a familiar urban environment is by means of mental maps: a memory representation of spatial information in our minds. A highly *imageable* urban environment has a high probability of evoking a strong image among various observers. This would be one whose "districts", "nodes", "landmarks" or "pathways" are easily identifiable and are easily grouped into an overall coherent pattern.

The mental map is primarily the representation of spatial relationships and contains some map-like qualities[2]. They consist of the topological relationships of spaces rather than their absolute coordinates and distances, which often bear little resemblance to the real environment. This is because our mind tends to simplify visual patterns by reducing complex spaces to a simple collection of basic shapes[3]. Also, the mental map blurs distance and direction but treats the topological relations with great clarity[4]. People's spatial knowledge is a complex collection of varyingly perceived items, qualities, and events, which is in effect a multi-modal representation of the city[5].

However, the relationships between the individual memory and physical elements of the built environment are rarely analyzed quantitatively. This is largely due to the insufficient relevant datasets. The employed data collection methodologies are frequently limited to manual ones such as hand-drawn sketch maps, interviews, and questionnaires. Although sketch maps provide some general spatial cognition measurements such as the relative location of places, their shapes, or even perceived distances between places, the results of sketch maps are very hard to quantify and compare. In addition, sketch maps may vary as a result differences in individuals' drawing abilities. Supplementing maps with interviews requires a considerable amount of time to interact with participants and the translation of images into words may be unreliable.

This study takes a different approach by applying a web-survey for data collection, measuring the memories of participants. The proposed data collection technique enables us to externalize participants memories and assemble larger-scaled datasets. The question is whether better spatial knowledge is a function of the legibility of the city and of temporal factors, particularly whether spatial knowledge depends upon the time spent in a place. We try to answer this question by exploring whether spatial comprehension is influenced by the pattern of the city and individuals' daily activities. Both of these are examined considering temporal factors -- how the outcomes change over time.

## 2. Analytical framework

One major challenge that researchers face when investigating people's mental images of the city is how to externalize an individual's spatial knowledge of a familiar environment. Although several kinds of analytical frameworks are proposed, the previous research can be classified into two groups.

The first group employs traditional manual based data collection techniques to obtain qualitative data[1,2,5,6]. The most common practice is to have subjects produce hand-drawn sketch maps of a specific urban area based on their memories. Other commonly used methods are verbal interviews, questionnaires, and cognition tasks. The second group relies on the recent development of computational technologies (e.g., crowdsourcing) together with the quantitative analysis, which enable us to study directly the relationship between the physical appearance of cities and human behavior[7,8,9,10,11,12]. Some previous studies in this group deal with the memorability of visual information[13,14,15]. "Memorability" has been measured by analyzing photographs derived from a web-gaming platform, which presents a series of unique photos, with repeat photos interspersed, making it possible to quantify systematically how people remember images.

These methods enable estimates to be made of the collective sensuous of a city, but may not be adequate to capture spatial knowledge or how it develops over time. Since people's spatial knowledge and the memory of the place are largely acquired through repeated encounters of the environment and observer, disregarding the urban structure and habits of use makes the data unreliable. Finally, any differences in the subject's accumulated knowledge and potential biases in perception are typically not considered. For example, a subject's spatial knowledge is partly, a function of the number of visits they have made to a place, indicating that it is also dependent on a temporal factor. Without being able to control for these factors and potential biases, it is difficult to use the dataset for a reliable analysis the image of city.

This paper evaluates inhabitants' familiarity with a neighborhood through web-based visual surveys in the form of a guessing game. By using online visual surveys, the availability of participants has exponentially increased. It also enables us to capture detail in a comprehensive way about every corner of the urban environment. The methodology avoids conducting lengthy interviews with selected subjects, while the carefully designed user interface of the website makes it possible to perform required actions without face-to-face guidance.

## 3. Settings for data collection and collected datasets

We designed a web-based visual survey- Urbanexplorer- as a game to help people explore their familiar urban neighborhood. Urbanexplorer takes advantage of thousands of geo-tagged streetview images retrieved from the Google streetview API.

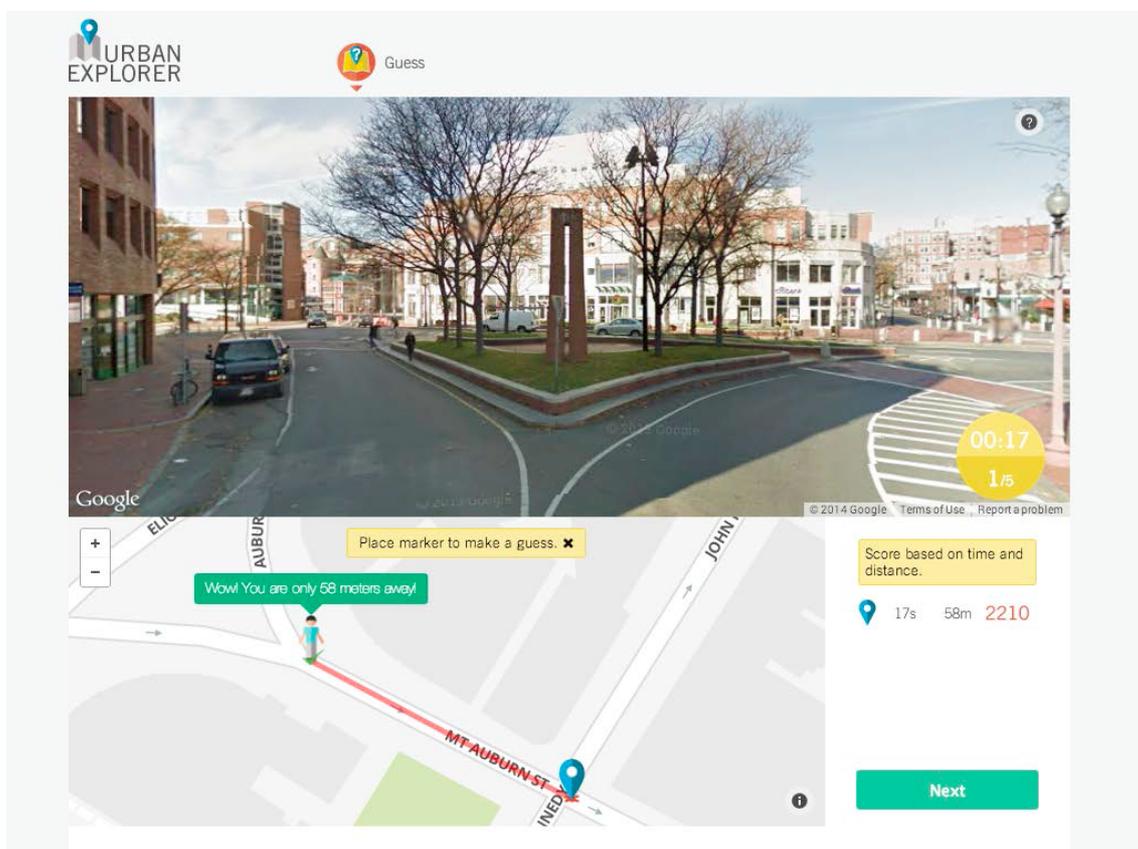

**Figure 1.** The user interface of Urbanexplorer

Figure 1 shows the user interface of UrbanExplorer. Participants were shown random streetviews from a neighborhood they were familiar with and were asked to guess where this place was by pointing to a map. Guesses were scored based on the time taken to place the guess, along with the distance between the guessed location and the real location of the image. This indicates that each individual's familiarity with a given place is a function of reaction time and guessed distance to answer. The overall familiarity of the place can be inferred from the collective actions and responses of all players.

The urban area chosen for this study is Harvard Square in Cambridge, MA. Adjacent to Harvard University, Harvard Square is an area frequently visited by tourists and local residents. It consists of a variety of local attractions, such as restaurants, salons, clothing shops, bookstores, coffee shops, florists, and hotels. The Harvard Square MBTA stop is a major transportation hub that links the subway, buses, and taxis, transporting thousands of people to and from this area every day. The road system of this area is complex and irregular. Most of the roads don't follow a grid, and most of the major intersections are not perpendicular. The highly visited public places, various land users, and irregular structure renders it an ideal location for our study.

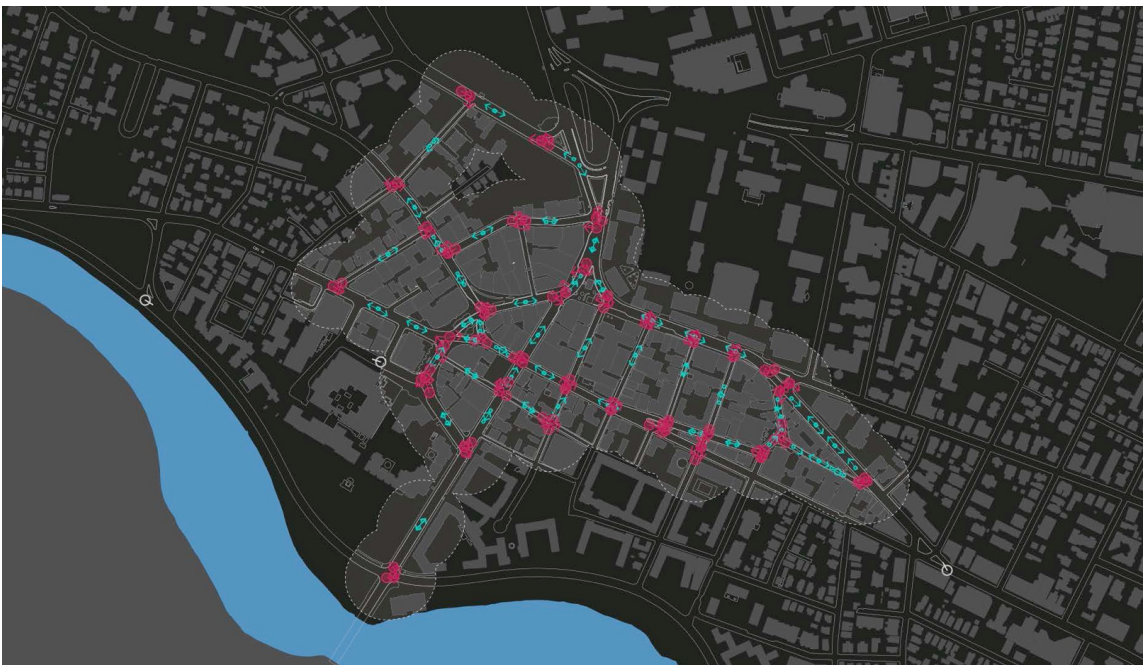

**Figure 2.** Location of all street views selected around Harvard Square

The images around Harvard Square were categorized into two groups representing nodes and links. According to Lynch's five elements[1] and Golledge's anchor point theory[6], the nodes are images of intersections, public squares, and single buildings that are usually referred to as spatial reference points or landmarks, and the links are images of the street sections between the nodes. Golledge[16] argued that landmarks acted as anchor points for organizing other spatial information into a pattern. The complex urban forms stored in our memory in the form of a linked-node configuration. The process of acquiring spatial knowledge involves continuously adding new nodes to the existing node-link framework[17]. For each node, 3 to 4 images are selected looking at different points of view. For each link, 2 images are selected looking at different directions.

| Stored information | Description of information |
|---|---|
| Place ID | Each streetview is given an unique ID for easy retrieving |
| Time spent to guess | Time is calculated from the moment a streetview is shown to the moment the person clicks the guess button |
| Location of streetview | Longitude and latitude of the streetview |
| Location of guess | Longitude and latitude of the guessed location |
| Distance to answer | Distance is calculated between the two locations above |
| User information | Information of the user matched with his/her survey |
| Time of creation | Time when the guess is made |
| Score | Score is a function of distance and time spent based on a scale of 1 to 2500 |

**Table 1.** Attributes associated with each guess

Collecting data on the demographics of the participants, the website asks them to answer a three-question survey after they placed three guesses looking at images of Harvard Square. The survey questions are: (1) How do you know Cambridege? (2) How old are you? (3) How many times have you visited Harvard Square? Their replies to these questions are used to filter for participants who have never been to Cambridge or the studied area. All the guesses were then stored in the database with a set of attributes (Table 1).

In total, we collected samples from 394 people for this paper. A total of, 4216 guesses were made, of which 3617 were made by people who filled out the survey. Most of participants were students or faculty members, and were people who currently live and/or work in Cambridge. Among them, 68.0% of the participants stated they currently live or work in Cambridge or have done so in the past; 79.4% of the subjects claimed to have visited the studied area more than 10 times in the past; 15.5% claimed to have visited the studied area 3-10 times; 84.0% of the subjects are under 35 years old.

### 4. Results

**Most familiar and least familiar places**
The places with the highest scores are primarily public spaces at the center of the area, such as Harvard Square, Brattle Square, and Winthrop Square. Certain stores at the center of Harvard Square (e.g., Harvard Coop) along with Massachusetts Avenue also received high scores. In fact, the place that is most recognized is the Qdoba Mexican Grill along Massachusetts Avenue, with a score of 0.847. The links with the highest score are streets within one block to the center (e.g., JFK Street and Massachusetts Avenue). This finding indicates that the degree of interaction and the proximity to the center is highly correlated with the probability of being recognized by most participants. The observation also suggests that the irregularity of urban structure doesn't prevent the formation of strong mental images. Most of the highly-recognized places are in the center of the studied area, where road system is extremely complex. The shapes of the highly-recognized squares are triangular and the roads travelling through them are curved. This finding seems to be contradict to our intuition that a clear urban structure is more likely to evoke strong mental images. However, a legible urban structure doesn't always and necessarily indicate a regular city grid. Rather, patterns of use trump geometric regularity, and even distorted streets can have a high level of imageability[1].

The places with the lowest scores are primarily school or residential buildings with no eye-catching signs or distinct features, as well as streets that don't have much activity (e.g., Garden Street, Bow Street and Arrow Street). The least recognized places are all located away from the center of the studied area. Some findings that are worth mention include:

(1) The second least recognized streetview with a score of 0.022 is the building of the John F. Kennedy School, located at the intersection of JFK Street and Eliot Street. JFK Street is the busiest street in Harvard Square with heavy traffic and frequent congestion. The fact that the building at one busy intersection is very unfamiliar to people indicates that frequently passing by a place does not necessarily evoke strong images. We speculate that institutional areas engage fewer of the passersby thank commercial areas.

(2) The sixth least recognized streetview with a score of 0.074 looks at the back of the Harvard University Office Building (Smith Campus Center), which takes up the entire block. However, the eleventh most recognized streetviews looks at the front of the same building, where there are shops and large red windows dominate the three-story podium in the center of the view. This observation suggests that tall structures in a dense environment do not necessarily evoke strong mental images, because pedestrians are more likely to look at features visible at eye level.

**Links and Nodes**
Of all the answers placed within 100 meters of the correct location, 30% of the answers of the nodes were concentrated within 20 meters of the true location, whereas only 15% of the answers to the links were within 20 meters. This indicates that if people recognize the image, they tend to place the marker more accurately (within 20 meters to the real location) when the image they are guessing is a node. On the other hand, guesses for links tend to be less accurate.

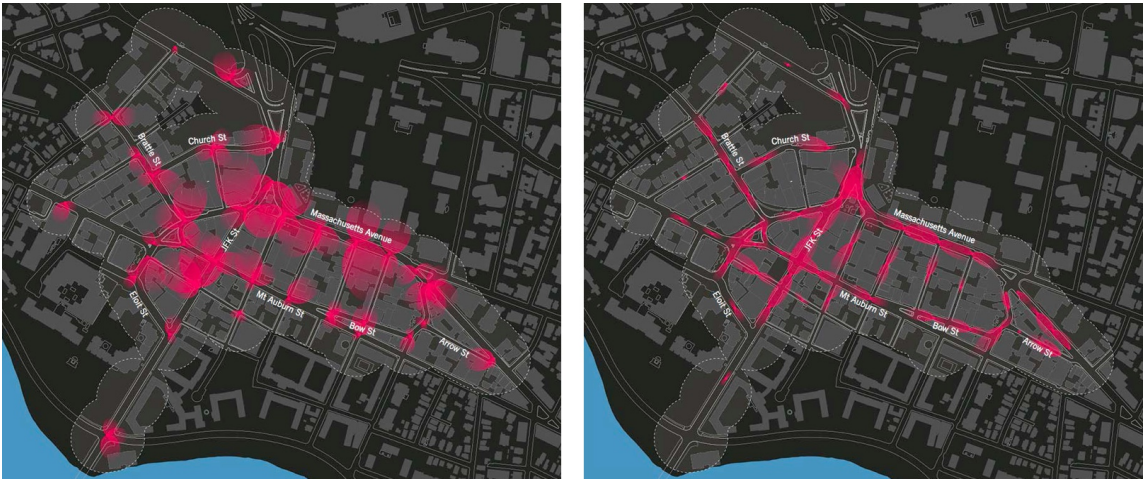

**Figure 3. (a)** Score of nodes.  **(b)** Score of links

Figure 3 (a) shows the scores of all nodes. The size of the ellipses indicates the score. The distribution of the node's scores are mostly coincided with our observation of the pedestrian traffic patterns. The nodes more familiar to people are located along Massachusetts Avenue and JFK Street, which are the two busiest streets. Most of the nodes along these two streets point to stores or shops along the street. Along Massachusetts Avenue, the more familiar views are mostly towards the south side of the street where shops predominate, while the North Side is mainly institutional buildings. Likewise, among the nodes along Mt. Auburn Street, the more familiar views are all towards the south. These two parallel streets seem to act as a magnet that draws more attention to the area due to the presence of many restaurants and cafes.

Figure 3 (b) shows the scores of all links. The more familiar streets are Massachusetts Avenue and JFK Street. A pattern can be observed among the streets between Massachusetts Avenue and Mt. Auburn Street. The closer the street is to the center, the more familiar it is to most people. This again verifies the earlier observation that proximity to center is strongly correlated with the degree of familiarity.

An interesting finding was observed when conducting high-resolution analysis of every angle of this urban area. This experiment asked participants to guess streetviews without rotating the view, which allowed us to compare the familiarities of different viewpoints at the same geographic location. At most locations, the familiarities of different viewpoints vary to some extent. In fact, at certain locations, the familiarities between different viewpoints are dramatically different. For example, the shop next to the Qdoba Mexican Grill- the most recognized place- is very poorly recognized, although these two places are only 10 meters apart. At viewpoint A, the Qdoba Mexican Grill, almost 70% of participants made a guess within 20 meters. In contrast, at viewpoint B, Zinnia Jewelry, only 5% of participants made a guess within 20 meters, and almost 80% of participants didn't even guess within 800 meters, which suggested that they couldn't recognize this place at all. We speculate that Qdoba is a fast food restaurant, which people frequently visit, whereas Zinnia is a small jewelry shop that does not attract many people. Also, we speculate that the church tower at the very far end of image A provides a hint to the location of this streetview, suggesting that it works as the anchor point, as a spatial reference.

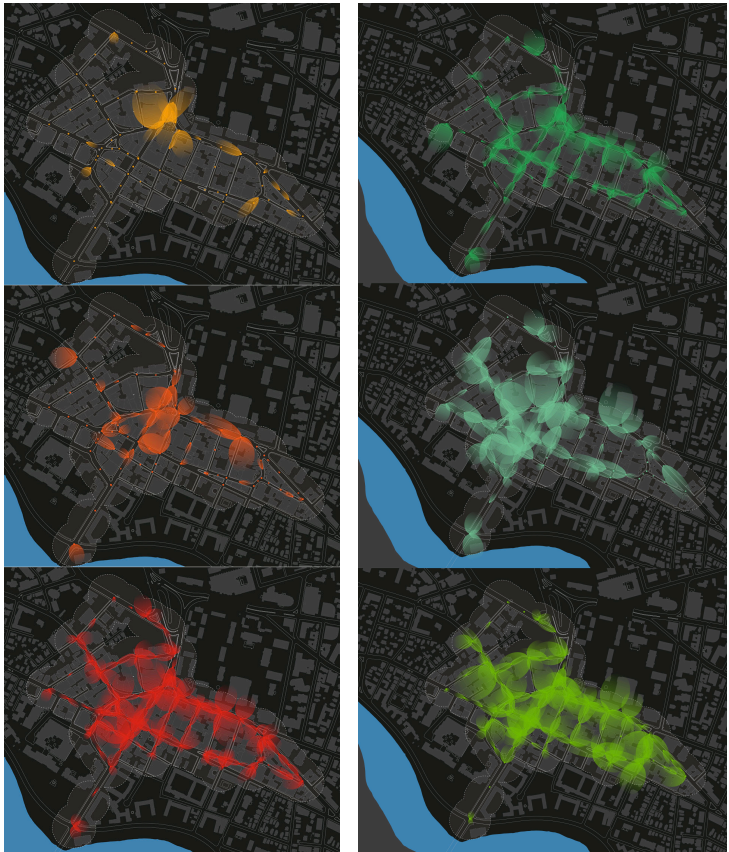

**(a)** **(b)**
**Figure 4. (a)** above: score of people who visited less than 3 times. Middle: score of people who visited 3-10 times. Below: score of people who visited than 10 times. **(b)** above: score of people who live in Cambridge. Middle: score of people who work in Cambridge. Below: score of people who live and work in Cambridge.

Figure 4 (a) shows how spatial knowledge differs depending upon the number of visits. From top to bottom, images show the familiarity of places among people who have visited this area less than three times, between three and ten times, and more than ten times. When people visit once or twice, spatial knowledge of the activity center is first acquired. When more visits are paid, people start to gain knowledge of the spaces that are spatially close to the center, including the part of the streets that connect to the center of this area, along with intersections that are one or two blocks away from the center. Finally, when people have visited the place more than ten times, knowledge of the spatial structure starts to appear. This observation seems to favor the anchor-point theory, which proposes that the spaces in our mind are formed as a linked-node structure. However, the anchor-point theory is mainly concerned with each individual's spatial knowledge. It might not apply to the collective mental maps of the city. Our data suggests that spatial understanding tends to grow outward from the anchor points.

Figure 4 (b) shows how modes of interaction with the city affect spatial familiarity. From top to bottom, images show the familiarity of places among people who only live in Cambridge, who only work in Cambridge, and who both live and work in

Cambridge. The familiarities among people who only live in Cambridge extend more towards the surrounding area, where there are mostly residential districts, as opposed to being concentrated in the center activity area. In comparison, the familiarities among people who only work in Cambridge concentrate along the busy streets and center activity area and not so much on the parallel streets in between Massachusetts Avenue and Mt Auburn Street. The familiarities among people who both work and live in Cambridge are confined to the middle portion of the studied area, framed by Massachusetts Avenue and Mt Auburn Street, including the parallel streets in between. All the places in the middle area are very well recognized by people who live and work in Cambridge. The observation suggests that the modes of interactions can affect the distribution of familiar places. In general, people know places better if they both work and live in the area.

## 5. Discussion

Results indicate that the places closer to the urban center is more likely to be remembered. We speculate that people's strong image of the city is formed along with the familiar paths or nodes, which are the nucleus in the district. Then, their spatial knowledge expands outward, which corresponds to Lynch's hypothesis[1]. In addition, we suggest that inhabitants' memory of the city largely depends on patterns of their daily activities and their resident status in the city. People, who are both living and working in the city tend to increase the quality of their spatial knowledge better than the ones who either only live or work there. Thus, the degree of interaction determines the level of familiarity, and modes of interaction affect distribution of familiar places. Also, our finding suggests that large-sized and taller buildings do not necessarily evoke strong mental images because pedestrians are more likely to look at features visible at eye level. The distinctive features of a building can evoke high imageability if the feature is constantly visible from a pedestrian's viewpoint, however, places along a busy street may have poor recognition if most of the passersby have limited or no interaction with it.

These findings are largely consistent with previous research, which has revealed that groups with different activity patterns tend to produce disparate mental images. Also, the degree of local understanding of places largely depends on the length of stay in the city[6]. There are correlations between the degree of familiarity, the length of residency, the location for job and dwelling, and their attributes[2].

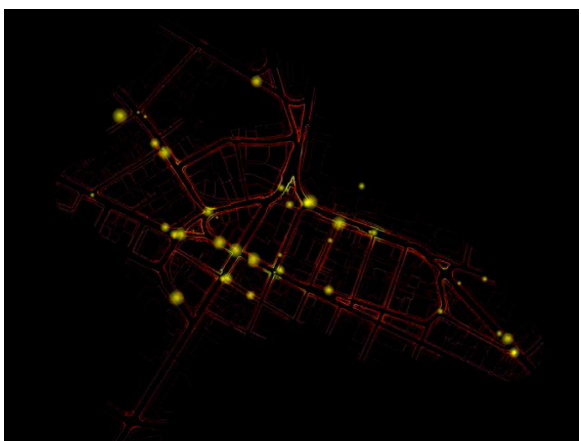
(a)

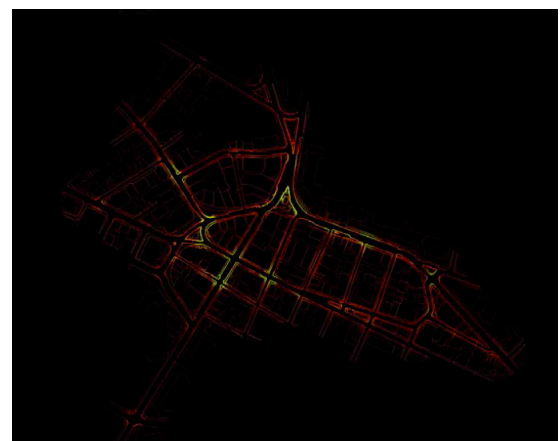
(b)

**Figure 5. (a)** the collective memory of the city from view point of the node. **(b)** the collective memory of the city from view point of the link.

On the other hand, our results also diverge from previous studies. First, we discovered that the irregularity of urban structure doesn't prevent the formation of strong mental images shown in Figure 5. It presents the collective memory of the city using people's preferences, highlighting the most remembered and least remembered sections of the cityscape.

Lynch (1960)[1] argued that the highly "legible" urban structure can be helpful to increase people's imageability. However, he also argued that such imageability derives not only from the designed and formalized structure (such as a regular grid), but also from more subjective and "fuzzy" perception and memory related to a heuristic visual perception of a place over time. This indicates that the distorted street is not always the cause of the visual chaos, but can be the spatial reference that provides the subject with a mode to identify a place and orient themselves at the neighborhood scale. Although our findings seem to contradict the popular interpretation of Lynch (e.g., that legible structure requires geometric regularity), they confirm that a visual hierarchy of the streets can be the result of use patterns even in a complex pattern of streets.

Second, our analysis shows that the different viewpoints of the same location might induce either extremely high or low familiarity. Views of the same building from different angles may also reduce familiarity: the back view of an office building along a busy street is the 6th least remembered view along 190 streetviews in that area, whereas the front view of the building is the 11th most remembered view. The use of Google's geo-tagged Street Views for crowd-sourced data collection enables us to conduct this high resolution analysis, which was not possible prior to this study.

Finally, we discovered that the frequency of visits to the district directly correlated with the degree of the richness of spatial knowledge. With limited visits, people are only able to remember some spots that they have limited interactions with. The more visits paid, the more spots will be added to their memory reaching outward from the vicinity of the center. This, to some extent, aligns with Golledge's anchor-point theory[6], indicating that spatial knowledge was acquired over time.

All of these facts suggest that perceptual knowledge and the process of accumulating it have a strong correlation with the spatial layout of the built environment. Human activity patterns are the drivers of spatial knowledge, which in turn largely depends on temporal parameters.

The research method we employed provides value and novel perspectives to the subject of imageability, but it also has several limitations. One concern is the lack of control over the demographics of the participants. People who are willing to take the survey are more likely to be medium or higher income groups, who have access to the Internet and are willing to voluntarily participate in academic research. Another problem is that there is no means to filter irrelevant data, including the readability of images. People might carelessly click around without paying attention to the instructions and there is a certain bias deriving from the ability to interpret a given image among individuals. Future improvements can be made using machine learning to control and balance demographics of the users, and designing a better user interface to eliminate user errors. Nonetheless, the application of the web-based visual survey in the form of a geo-guessing game allows researchers to rapidly elicit spatial knowledge among a large number of city dwellers, and allows researchers to conduct quantitative data analysis.

These findings are useful for architects and urban planners in revealing potentially highly imageable locations of the city, and can provide guidance for designing vivid urban environments that facilitate rapid learning and easy orientation[18]. The tool can be adapted to researching the spatial knowledge acquisition in an unfamiliar neighborhood, cognitive representation of spatial relationships, and results of perceptional knowledge on human mobility behavior. This is a piece of critical information that was not possible to obtain prior to this study.